\documentclass[useAMS,usenatbib]{mn2e}
\usepackage{graphicx,fleqn,natbib,amssymb}
\DeclareGraphicsExtensions{.eps,.cps,.ps,.eps.gz,.ps.gz,.eps.Z}
\usepackage{subfigure}
\usepackage{multirow}

\usepackage{graphicx}
\usepackage{epstopdf}
\usepackage{threeparttable}

\newcommand{\comments}[1]{}

\title[Type IIb SN 2009mg]{Multi-wavelength Observations of the Type IIb Supernova 2009mg\thanks{This paper includes data 
gathered with the 6.5-m Magellan Telescopes located at Las Campanas Observatory, Chile; and the Gemini Observatory, 
Cerro Pachon, Chile (Gemini Programme GS-2009B$-$Q$-$9, GS-2009B$-$Q$-$40 and GS-2009B$-$Q$-$67).}}

\author[Oates et al.]{S. R. Oates$^{1}$, 
A. J. Bayless$^2$, 
M. D. Stritzinger$^{3,4,5}$
T. Prichard$^{2,6}$, 
J. L. Prieto$^{7}$,
\newauthor S. Immler$^{8,9,10}$, P. J. Brown$^{11}$, A. A. Breeveld$^{1}$,  M. De Pasquale$^{12}$, N. P. M. Kuin$^{1}$, 
\newauthor  M. Hamuy$^{13}$, S. T Holland$^{14}$, F. Taddia$^{4}$, P. W. A. Roming$^{2,6}$\\
$^{1}$ Mullard Space Science Laboratory, University College London, Holmbury St. Mary, Dorking Surrey, RH5 6NT, UK; sro@mssl.ucl.ac.uk \\
$^{2}$ Southwest Research Institute, Space Science and Engineering, Division 15, Bldg. 263, 6220 Culebra Rd., San Antonio, TX 78238 \\
$^{3}$ Department of Physics and Astronomy, Aarhus University, Ny Munkegade 120, DK-8000 Aarhus C, Denmark \\
$^{4}$ Department of Astronomy, The Oskar Klein Centre, Department of Astronomy, Stockholm University, AlbaNova, 10691 Stockholm, Sweden \\
$^{5}$ Carnegie Observatories, Las Campanas Observatory, La Serena, Chile \\
$^{6}$ Department of Astronomy and Astrophysics, Pennsylvania State University, 104 Davey Laboratory, University Park, PA 16802 \\
$^{7}$ Department of Astrophysical Sciences, Princeton University, Peyton Hall, Princeton, NJ 08544, USA; Hubble, Carnegie-Princeton Fellow   \\
$^{8}$ Department of Astronomy, University of Maryland, College Park, MD 20742, USA \\
$^{9}$ Astrophysics Science Division, Code 660.1, NASA Goddard Space Flight Centre, 8800 Greenbelt Road, Greenbelt, Maryland 20771, USA \\
$^{10}$ Centre for Research and Exploration in Space Science and Technology Code 668.8 8800 Greenbelt Road Goddard Space Flight Centre, \\
   Greenbelt, MD 20771, USA \\
$^{11}$ Department of Physics \& Astronomy, University of Utah, 115 South 1400 East \#201, Salt Lake City, UT 84112, USA\\
$^{12}$ Department of Physics and Astronomy, University of Nevada, Las Vegas - 4505 S. Maryland Parkway, Las Vegas, NV 89154, USA\\
$^{13}$ Departamento de Astronom\'{i}a, Universidad de Chile, Santiago, Chile\\
$^{14}$ Space Telescope Science Institute,3700 San Martin Drive,Baltimore, MD 21218,USA\\
 }

\begin{document}

\date{Accepted...Received...}

\maketitle

\label{firstpage}

\begin{abstract} 
We present {\it Swift} UVOT and XRT observations, and visual wavelength spectroscopy 
of the Type IIb supernova (SN) 2009mg, discovered in the Sb galaxy ESO 121-G26. The 
observational properties of SN~2009mg are compared to the prototype Type IIb 
SNe 1993J and 2008ax, with which we find many similarities. However, minor 
differences are discernible including SN~2009mg not exhibiting an initial fast 
decline or $u$-band upturn as observed in the comparison objects, and its rise to 
maximum is somewhat slower leading to slightly broader light curves. The late-time 
temporal index of SN~2009mg, determined from 40 days post-explosion, is consistent with the 
decay rate of SN~1993J, but inconsistent with the decay of $^{56}$Co. This suggests leakage 
of $\gamma$-rays out of the ejecta and a stellar mass on the small side of the mass 
distribution. Our XRT non-detection provides an upper limit on the mass-loss rate of the 
progenitor of $\dot{M}<1.5\times 10^{-5}\; M_{\odot} \; {\rm yr^{-1}}$. Modelling of 
the SN light curve indicates a kinetic energy of $0.15^{+0.02}_{-0.13}\times 10^{51}\;{\rm erg}$, 
an ejecta mass of $0.56^{+0.10}_{-0.26}M_{\sun}$ and a $^{56}$Ni mass of $0.10\pm0.01M_{\sun}$.

\end{abstract}


\section{INTRODUCTION}
\label{intro}

\begin{figure*}
\includegraphics[angle=0,scale=0.2]{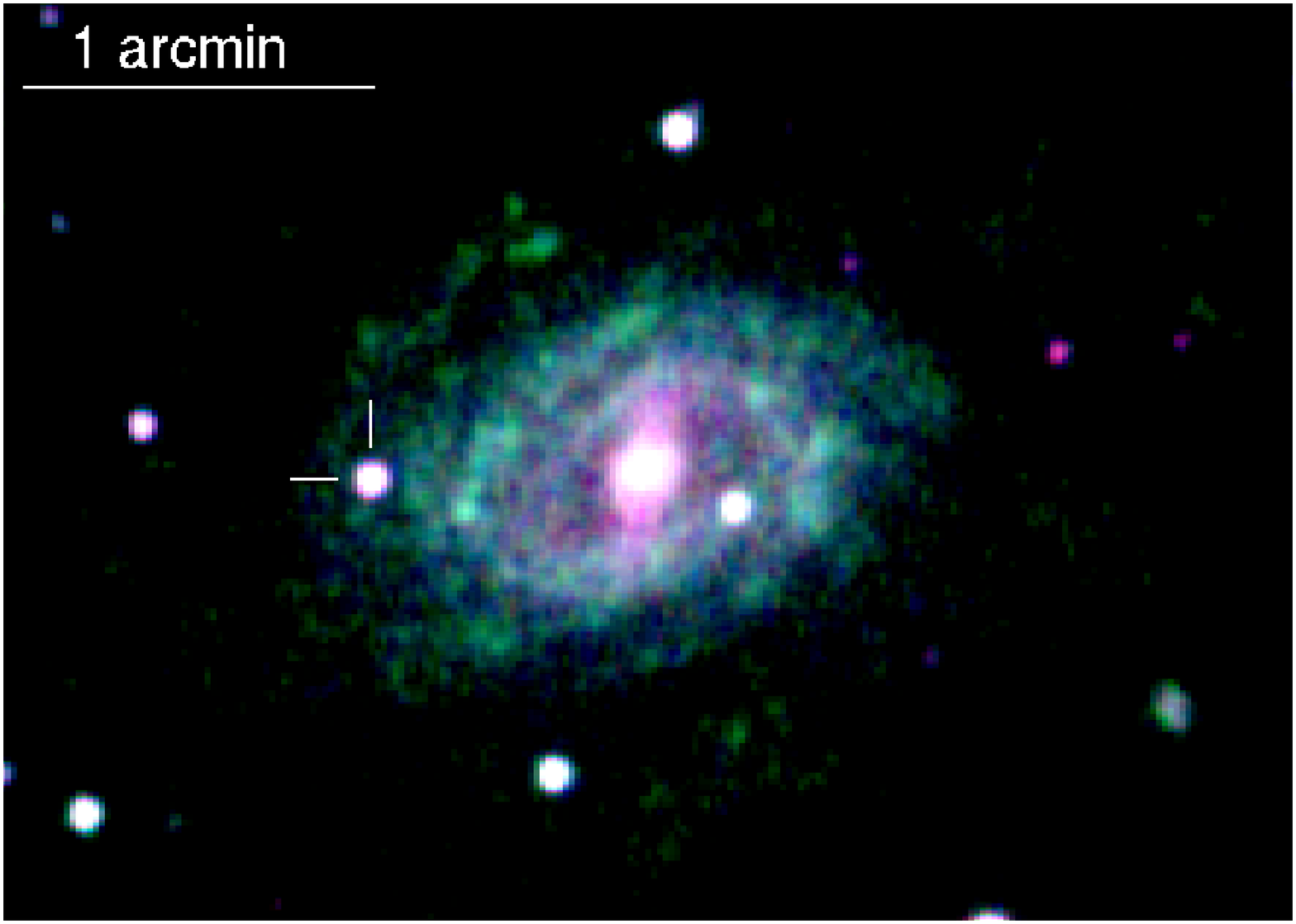}
\includegraphics[angle=0,scale=0.2]{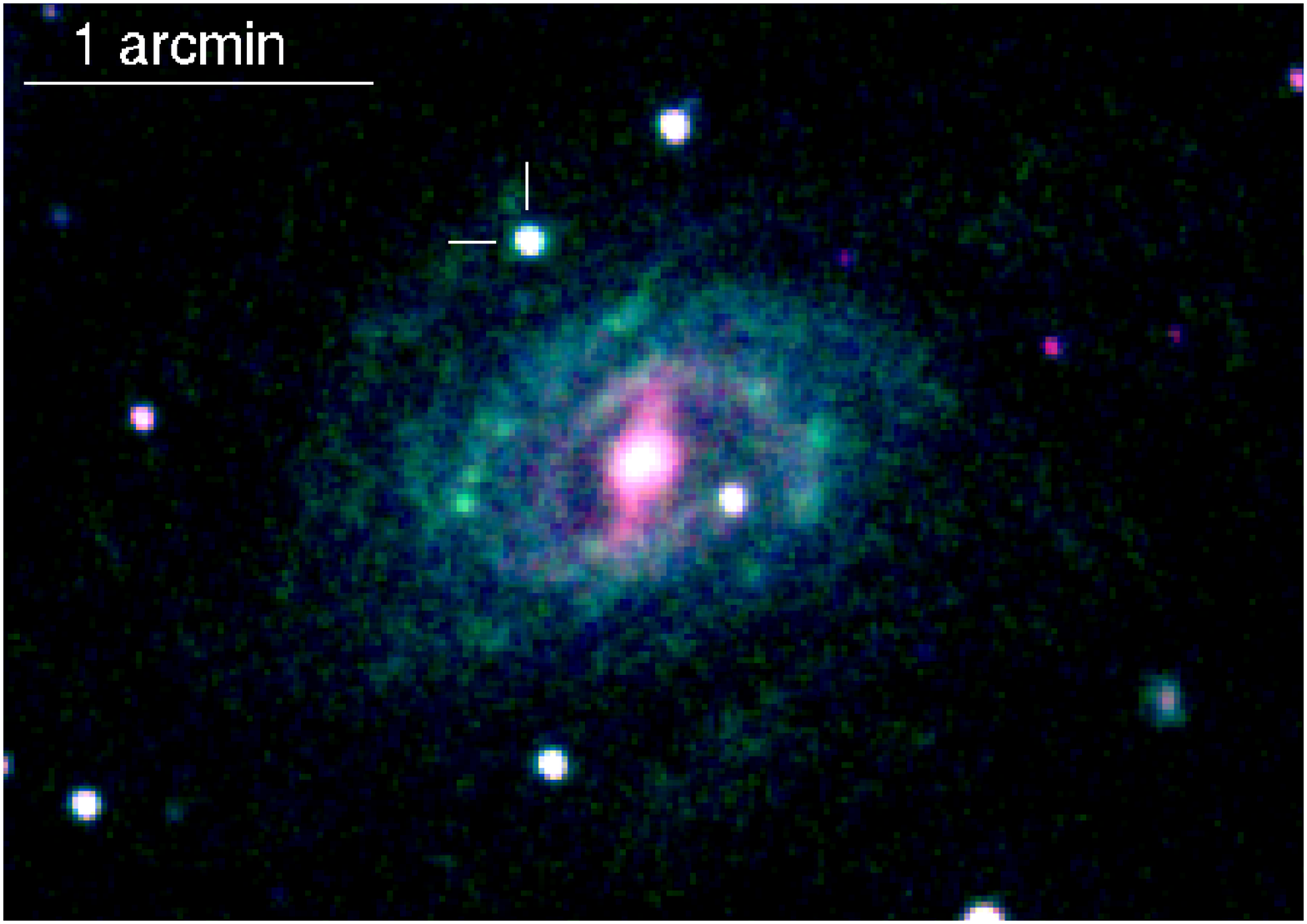}
\caption{Two panels displaying the {\it Swift} UVOT images of the galaxy ESO 121-G26 and SNe 2009mg ({\em left})
and 2008M ({\em right}), respectively. The RGB images were constructed from $u$, $b$, and $v$ images.}
\label{optimage}
\end{figure*}

Core collapse supernovae (SNe) are the death throws of massive stars. 
These stellar explosions are classified into 
Type II and Type Ib/c sub-types depending on their spectroscopic 
characteristics. The spectra of SNe~II
contain prevalent hydrogen lines, while those of SNe~Ib lack hydrogen lines, 
but display lines of helium. 
The spectra of SNe~Ic, on the other hand, lack both hydrogen and helium lines 
\citep[see][for further details and references]{fil97}. 
The lack of hydrogen lines in SNe~Ib/c is thought 
to be due to the hydrogen envelope being stripped before the final 
pre-SNe stage is reached. 
In the case of SNe~Ic, the helium envelope is also thought to be stripped or 
$^{56}$Ni is not sufficiently mixed into any helium layer and thereby prevents 
the excitation of helium lines. The progenitors of SNe~Ib/c are therefore thought 
to be stars in which the majority of their hydrogen-rich envelope is stripped 
via  stellar winds and/or transferred to a secondary star via Roche-lobe over 
flow \citep[e.g][]{whe85,woo93,woo95}.

In 1987 hints appeared of a new sub-class of SNe~II from the spectral evolution 
of the peculiar SN~1987K \citep{fil88}. Around maximum light SN~1987K exhibited 
spectral characteristics reminiscent of a normal SN~II event including the presence 
of a broad H$\alpha$ absorption feature. However nearly five months later, when 
the object reappeared from behind the sun, its spectrum resembled more closer that 
of a SN~Ib. This peculiar spectral metamorphosis provided a hint that normal 
SNe~II maybe linked to the hydrogen stripped SN~Ib subclass. Similarly, the 
well-observed SN~1993J exhibited a clear SN~II-like spectrum just after 
explosion, but over several weeks its spectrum evolved to resemble that 
of a classic SN~Ib event with dominant He~I features \citep{fil93,swa93}. 
This led to the introduction of the Type~IIb subclass \citep{fil88}. 
SNe~IIb progenitors are able to retain a small amount ($\sim$ 0.01 M$_{\sun}$) 
of hydrogen at the time of explosion, and this residual hydrogen shell is 
manifested in the optical spectrum obtained just after explosion \citep{pod93,nom93}. 
Recently, it has been proposed that SNe~IIb may be further classified 
in to eIIb and cIIb sub-classes, dividing SNe~IIb into extended or compact progenitors, 
respectively \citep{che10}. This subclassification depends on the radius and 
mass-loss history of the progenitor star, which may be determined from X-ray 
and radio observations. Examples of SNe~eIIb are SN1993J and 2001dg, while examples 
of SNe~cIIb are 1996cb, 2001ig and 2008ax \citep{che10}.

Over the past 20 years, approximately 
72\footnote[1]{http://heasarc.gsfc.nasa.gov/W3Browse/star-catalog/asiagosn.html} \citep{bab99} 
SNe~IIb have been discovered, but only a handful have been well observed 
including, amongst others, SNe~1987K \citep{fil88}, 
1993J \citep{sch93}, 1996cb \citep{qiu99}, 
2003bg \citep{ham09}, 2008ax \citep{pas08,rom09,tau11,cho11} and most 
recently 2011dh \citep{arc11,mau11,sod11}. 
The exact number of SNe~IIb is uncertain due to the evolution of their spectra. 
If spectral observations of SNe~IIb are not obtained early enough then we may miss 
SNe~II-like features, such as prevalent hydrogen lines, 
and so misidentify SNe~IIb as SNe~Ib. 
Conversely, if late-time spectral observations are not obtained or are 
too poor for line identification, then the development of SN~Ib-like features, 
such as numerous He~I lines, may not be detected and so the 
SN could be misidentified as a SN~II. Currently, 
little is known about the spread in mass and energy of SNe~IIb, 
the exact manner of mass-loss, nor exactly how they relate SNe~II and SNe~Ib/c. 
It is important that an expanded sample of well-observed SNe~IIb is constructed 
as these objects with their small hydrogen envelope may bridge the gap between 
normal SNe~II and SNe~Ib/c. 

Direct information can be gathered on the progenitors of SN from pre-explosion images, 
but it is difficult to identify candidate progenitors without high resolution 
HST images. Candidate progenitor stars have been directly 
detected in pre-explosion images of the Type IIb SNe 1993J and 2011dh. 
In the case of SN~1993J, a K-type supergiant in a binary system was determined
to be its progenitor \citep{mau04}.
For SN~2011dh, a mid F-type yellow supergiant is observed at the location of 
the SN \citep{mau11,van11}, but it is currently too soon to determine 
whether this is the star that exploded, the binary companion, or an unrelated star 
\citep{arc11,sod11,bie12}. Evidently, further observations are required after SN~2011dh has 
faded to determine if the candidate star has disappeared or is indeed still present. 

\begin{figure*}
\includegraphics[angle=0,scale=0.70]{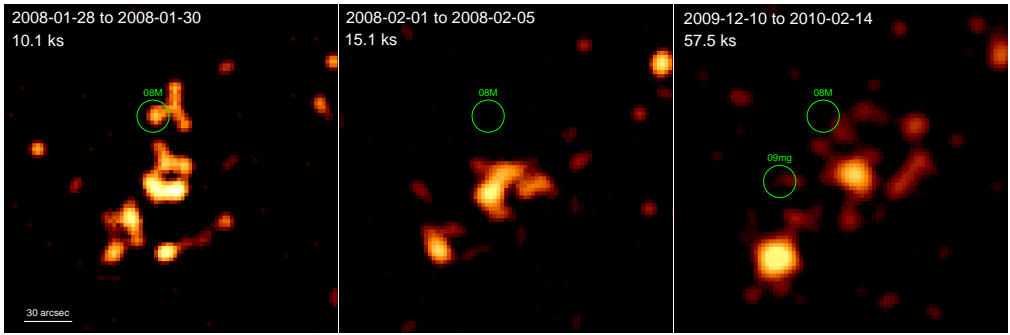}
\caption{The three panels display the {\it Swift} XRT images of the galaxy ESO 121-G26 at three different epochs. The
first panel consists of data from a two day period displaying the X-ray emission from SN~2008M. The middle panel consists
of data from a four day period, and shows that the X-ray emission from SN~2008M has faded. The final panel shows the position
of SN~2008M and SN~2009mg. No X-ray emission is detected from SN~2009mg.}
\label{xrtimage}
\end{figure*}

\subsection{Supernova 2009mg}
  
In this paper, we present broad-band UV and optical light curves and visual-wavelength 
spectroscopy of the Type~IIb SN~2009mg. 
SN~2009mg was discovered on the 7.9 December 2009 UT \citep{mon09} in the Sb galaxy ESO 121-G26.
With J2000 coordinates of RA $=$ $6^{h}21^{m}44.86^{s}$ Dec $=$ $-59^{o}44{'}26{"}$, 
SN~2009mg was 46$\arcsec$ east and 4$\arcsec$ south from the center of the host galaxy. 
The redshift of the host galaxy is $z=0.0076\pm0.0001$ \citep{kor04}. The distance 
to the host galaxy is therefore 32.7~Mpc and the distance modulus is $\mu=33.01\pm0.48$ mag, both of which were 
retrieved from the NASA/IPAC Extragalactic Database\footnote[2]{http://ned.ipac.caltech.edu/}.

As commonly found for SNe~IIb, the classification was continually revised as the 
object evolved. The initial classification
was that of a SN~Ia \citep{pri09}, however, soon after this 
it was changed to a broad-line SN~Ic \citep{rom09a}. Days later as the spectrum evolved it 
was clear this object was a SN~IIb \citep{str10}, as its early spectra strongly resembled that of 
the prototypical Type IIb SNe 1993J and 1996cb. 
SN~2009mg was not the first SN to be detected in ESO 121-G26. In the
previous year the Type II SN~2008M was also discovered in this galaxy \citep{gre08}.
The locations of both SNe~2009mg and
2008M are indicated in Fig.~\ref{optimage}.

The organization of this paper is as follows. In \S~\ref{obs} we provide 
the main observations of this SN, and describe the data reduction and analysis methods used. The main 
results are presented in \S~\ref{results}, while the discussion and conclusions 
follow in \S~\ref{discussion} and \S~\ref{conclusions}, respectively. 
Unless stated all uncertainties throughout this paper are quoted 
at 1$\sigma$. We have adopted the Hubble parameter 
$H_0\,=\,70$\,$\rm km\,s^{-1}\,Mpc^{-1}$ and density 
parameters $\Omega_\Lambda$\,=\,0.73 and $\Omega_m$\,=\,0.27.

\section{OBSERVATIONS}
\label{obs}

\subsection{{\em Swift} Imaging}

X-ray and optical/UV observations were performed simultaneously with the X-Ray 
Telescope \citep[XRT;][]{bur05} and the Ultra-Violet Optical Telescope \citep[UVOT;][]{rom00,rom04,roming}, 
the two narrow field instruments onboard {\it Swift} \citep{geh04}. Analysis of 57.5~ks 
of X-ray data, see Fig.~\ref{xrtimage}, does not reveal a source at the position of the 
SN. In Section~\ref{XRTresults} the XRT observations are used to place limits on the 
mass-loss rate of the progenitor.  

Optical/UV observations of both SNe~2008M and 2009mg were obtained with {\em Swift}.
SN~2008M was observed over 7 epochs from 28.9 January 2008 UT until 5.3 February 2008 UT. 
These images were obtained from the {\it Swift} archive and used to construct deep template 
images, which were then used to subtract away galaxy emission at the position of 
SN~2009mg in each of the science images.

{\em Swift} UVOT observations of SN~2009mg commenced on 10.9 December 2009 UT 
and concluded on 14.4 February 2010 UT. Twenty-two epochs of broad-band imaging was performed using three optical 
($u$, $b$ and $v$) and three UV ($uvw1$, $uvm2$ and $uvw2$) filters.

\begin{figure*}
\includegraphics[angle=-90,scale=0.50]{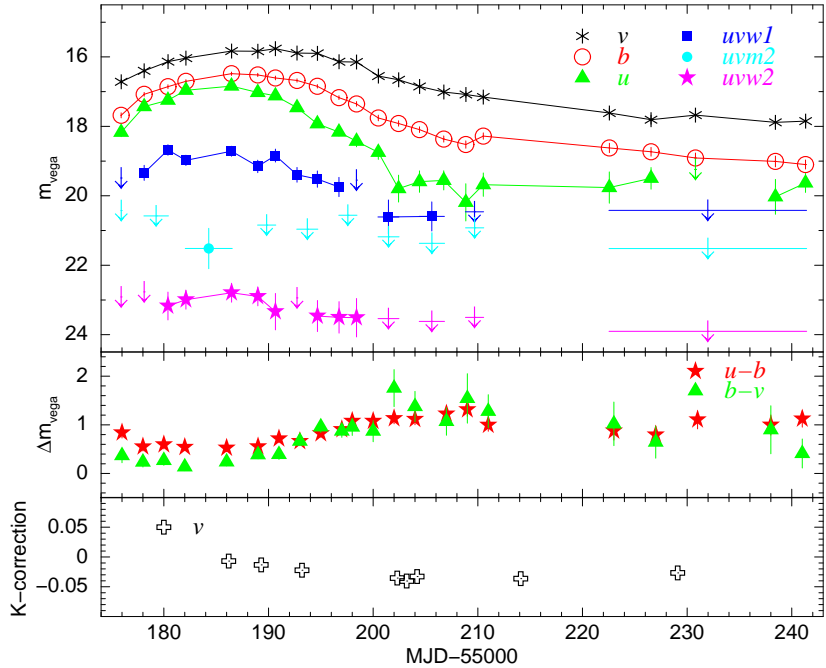}
\caption{{\em Upper} Panel: Observed UV and optical light curves of SN~2009mg.
{\em Middle Panel:} $(b-v)$ and $(u-b)$ colour curves, corrected for reddening.  
{\em Lower Panel:} K correction evolution for the UVOT $v$ filter during the course 
of UVOT observations.}
\label{UVOT_lc}
\end{figure*}

Photometry of SN~2009mg was computed following the method described in \citet{bro09}. 
The SN$+$galaxy photometry was extracted using a 3$\arcsec$ radius source region and 
a background region positioned away from the galaxy in a source free location. The 
galaxy count rate was determined from the summed exposures of SN~2008M, using the 
same background and source regions. This value was then subtracted from the SN+galaxy 
photometry determined from the images of SN~2009mg. The SN photometry was then aperture 
corrected to 5$\arcsec$ in order to be compatible with UVOT calibration. The analysis 
pipeline used software HEADAS 6.10 and UVOT calibration 20111031. Count rates were converted 
into magnitudes using the UVOT zero points presented by \cite{bre11}. The resulting light 
curves are presented in Fig.~\ref{UVOT_lc} and the photometry is provided in Table~\ref{UVOT_mag}. 
For some of the UV data points, several epochs have been coadded to provide detections or 
deep upper limits.

\subsection{Optical Spectroscopy}

Nine epochs of low-resolution and one epoch of high-resolution optical spectroscopy 
were obtained with facilities located at the Las Campanas Observatory and the Gemini-South 
telescope. A journal of the spectroscopic observations is provided in 
Table~2. 
Our spectroscopic series covers the flux evolution from 
$-$1.2 to $+$91.6 days relative to $B$-band maximum. 
All spectra were reduced in the standard manner using IRAF 
scripts\footnote{IRAF is distributed by the National Optical Astronomy Observatories, 
which are operated by the Association of Universities for Research in Astronomy, Inc., 
under cooperative agreement with the National Science Foundation.}, and in most cases, 
telluric corrections were not performed. 

\begin{table*}
\begin{tiny}
\begin{tabular}{|@{}l@{}|cccccc}
\hline
Time  & \multicolumn{6}{|c|}{---------------------------------------UVOT Observed Magnitudes---------------------------------------} \\
(MJD-55000+)   &        $v$       &        $b$       &       $u$        &    $uvw1$        &   $uvm2$         &    $uvw2$  \\
\hline
$175.92\pm0.03$  &  $16.67\pm0.07$  &  $17.68\pm0.07$  &  $18.17\pm0.13$  &  $>19.49$        &  $>20.42$        &  $>22.91$       \\
$178.13\pm0.04$  &  $16.40\pm0.06$  &  $17.08\pm0.06$  &  $17.43\pm0.08$  &  $19.34\pm0.22$  &  $  \cdots$      &  $>22.77$       \\
$179.26\pm1.17^*$  &  $ \cdots $      &  $ \cdots $      &  $\cdots$        &  $\cdots$        &  $>20.58$        &  $\cdots$        \\
$180.40\pm0.04$  &  $16.14\pm0.06$  &  $16.86\pm0.06$  &  $17.25\pm0.08$  &  $18.68\pm0.15$  &  $    \cdots  $  &  $23.18\pm0.40$ \\
$182.11\pm0.07$  &  $16.03\pm0.05$  &  $16.70\pm0.05$  &  $16.96\pm0.06$  &  $18.98\pm0.15$  &  $    \cdots  $  &  $22.99\pm0.27$ \\
$184.29\pm2.25^*$  &  $    \cdots   $  &  $    \cdots  $  &  $    \cdots$  &  $  \cdots    $  &  $21.52\pm0.58$  &  $\cdots$  \\
$186.47\pm0.07$  &  $15.83\pm0.05$  &  $16.48\pm0.05$  &  $16.84\pm0.07$  &  $18.72\pm0.15$  &  $    \cdots  $  &  $22.79\pm0.27$ \\
$188.98\pm0.04$  &  $15.76\pm0.05$  &  $16.52\pm0.05$  &  $17.02\pm0.07$  &  $19.15\pm0.18$  &  $    \cdots  $  &  $22.90\pm0.27$ \\
$189.82\pm0.88^*$  &  $    \cdots   $  &  $    \cdots  $  &  $\cdots      $  &  $    \cdots  $  &  $>20.84$        &  $\cdots$       \\
$190.66\pm0.04$  &  $15.76\pm0.06$  &  $16.61\pm0.06$  &  $17.12\pm0.09$  &  $18.84\pm0.19$  &  $   \cdots   $  &  $23.34\pm0.52$ \\
$192.73\pm0.03$  &  $15.89\pm0.05$  &  $16.68\pm0.05$  &  $17.46\pm0.08$  &  $19.39\pm0.21$  &  $   \cdots   $  &  $>22.94$        \\
$193.71\pm1.01^*$  &  $      \cdots $  &  $     \cdots $  &  $    \cdots$  &  $    \cdots  $  &  $>20.96$        &  $\cdots$         \\
$194.68\pm0.03$  &  $15.90\pm0.05$  &  $16.84\pm0.06$  &  $17.93\pm0.10$  &  $19.52\pm0.24$  &  $    \cdots  $  &  $23.46\pm0.45$ \\
$196.74\pm0.03$  &  $16.14\pm0.06$  &  $17.17\pm0.07$  &  $18.16\pm0.12$  &  $19.75\pm0.28$  &  $     \cdots $  &  $23.50\pm0.54$ \\
$197.58\pm0.87^*$  &  $     \cdots  $  &  $    \cdots  $  &  $    \cdots$  &  $     \cdots $  &  $>20.56$        &  $\cdots$         \\
$198.41\pm0.04$  &  $16.15\pm0.06$  &  $17.35\pm0.07$  &  $18.43\pm0.16$  &  $>19.55$        &  $    \cdots  $  &  $23.51\pm0.56$ \\
$200.50\pm0.04$  &  $16.55\pm0.08$  &  $17.76\pm0.08$  &  $18.75\pm0.20$  &  $     \cdots $  &  $    \cdots  $  &  $\cdots$        \\
$201.47\pm1.01^*$  &  $    \cdots   $  &  $    \cdots  $  &  $\cdots      $  &  $20.61\pm0.49$  &  $>21.18$        &  $>23.40$       \\
$202.44\pm0.04$  &  $16.67\pm0.07$  &  $17.92\pm0.08$  &  $19.79\pm0.35$  &  $     \cdots $  &  $     \cdots $  &  $\cdots$        \\
$204.44\pm0.03$  &  $16.85\pm0.07$  &  $18.10\pm0.08$  &  $19.59\pm0.30$  &  $     \cdots $  &  $     \cdots $  &  $\cdots$        \\
$205.60\pm1.21^*$  &  $     \cdots  $  &  $     \cdots $  &  $    \cdots$  &  $20.59\pm0.42$  &  $>21.37$        &  $>23.62$         \\
$206.76\pm0.07$  &  $17.01\pm0.07$  &  $18.36\pm0.09$  &  $19.56\pm0.26$  &  $     \cdots $  &  $     \cdots $  &  $\cdots$        \\
$208.86\pm0.04$  &  $17.08\pm0.08$  &  $18.52\pm0.11$  &  $20.19\pm0.54$  &  $     \cdots $  &  $     \cdots $  &  $\cdots$        \\
$209.69\pm0.87^*$  &  $     \cdots  $  &  $    \cdots  $  &  $    \cdots$  &  $>20.46$        &  $>20.92$        &  $>23.50$         \\
$210.53\pm0.04$  &  $17.16\pm0.08$  &  $18.28\pm0.09$  &  $19.68\pm0.34$  &  $    \cdots  $  &  $   \cdots   $  &  $\cdots$        \\
$222.57\pm0.04$  &  $17.62\pm0.13$  &  $18.62\pm0.13$  &  $19.76\pm0.45$  &  $    \cdots  $  &  $   \cdots   $  &  $\cdots$        \\
$226.56\pm0.07$  &  $17.81\pm0.13$  &  $18.73\pm0.13$  &  $19.50\pm0.31$  &  $   \cdots   $  &  $   \cdots   $  &  $   \cdots   $ \\
$230.80\pm0.11$  &  $17.68\pm0.12$  &  $18.91\pm0.14$  &  $>19.24$        &  $   \cdots   $  &  $   \cdots   $  &  $   \cdots   $ \\
$231.98\pm9.44^*$  &  $    \cdots   $  &  $    \cdots  $  &  $    \cdots$  &  $>20.42$        &  $>21.52$        &  $>23.90$         \\
$238.44\pm0.10$  &  $17.88\pm0.13$  &  $19.01\pm0.16$  &  $20.03\pm0.50$  &  $   \cdots   $  &  $   \cdots   $  &  $   \cdots   $ \\
$241.32\pm0.10$  &  $17.85\pm0.11$  &  $19.10\pm0.13$  &  $19.63\pm0.27$  &  $   \cdots   $  &  $   \cdots   $  &  $   \cdots   $ \\
\hline
${\rm m_{peak}}$   &  $15.75\pm0.05$  &  $16.46\pm0.04$  &  $16.78\pm0.07 $ &  $18.56\pm0.12$  & $\cdots$   & $22.50\pm0.23$ \\
${\rm t_{peak}}$   & $190.27\pm1.76$  & $187.40\pm1.11$  & $184.65\pm0.94$  & $183.82\pm3.64$  & $\cdots$  & $187.47\pm4.32$\\
${\rm M_{peak}}$   & $-17.68\pm0.48$  & $-17.10\pm0.48$  &  $-16.95\pm0.49$ &  $-15.30\pm0.50$ & $\cdots$  & $-11.37\pm0.53$ \\
\hline
\end{tabular}
\end{tiny}
\caption{Observed UVOT magnitudes and 3$\sigma$ upper limits, peak times (${\rm t_{peak}}$), peak magnitudes (${\rm m_{peak}}$), and absolute
peak magnitudes (${\rm M_{peak}}$). For some of the UV data points, several epochs have been coadded to provide 
detections and deeper upper limits. The resulting summed exposures are indicated with an $*$ in the time column. Absolute peak magnitudes 
include extinction and K corrections.}
\label{UVOT_mag}
\end{table*}

\begin{table*}
\begin{threeparttable}
\begin{tabular}{|@{}l@{}|@{}c@{}|@{}c@{}lccrcclc}
\hline
\multicolumn{2}{|c|}{Date of Observation} & Epoch\tnote{a} &~~~Telescope & Instrument & Spectral Range & Resolution & Exposure Time\\
(UT)        &(MJD-55000+)  &  (Days)  &  &  & (\AA) & (\AA) & (s) \\
\hline
21.2 Dec 2009 & 186.2 & $-$1.2   &~~~CLAY     & LDSS3 & 3900-9300  & 7 &2$\times$300 \\
24.3 Dec 2009 & 189.3 & $+$1.8   &~~~CLAY     & LDSS3 & 3900-9300  & 7 &2$\times$300 \\
24.3 Dec 2009 & 189.3 & $+$1.8   &~~~CLAY     & MIKE  & 3330-9150  & 0.15 & 1800 \\
28.2 Dec 2009 & 193.2 & $+$5.8   &~~~BAADE    & IMACS & 3976-6362;6489-10051& 9 & 600 \\
06.3 Jan 2010 & 202.3 & $+$14.8  &~~~GEMINI-S &  GMOS & 4201-8429  & 8 &1600 \\
07.2 Jan 2010 & 203.2 & $+$15.8  &~~~GEMINI-S &  GMOS & 4199-8426  & 8 &1800 \\
08.2 Jan 2010 & 204.2 & $+$16.8  &~~~CLAY     & LDSS3 & 3632-9516  & 7  & 600 \\
18.1 Jan 2010 & 214.1 & $+$26.6  &~~~GEMINI-S &  GMOS & 4201-8428  & 8 &1800 \\
02.1 Feb 2010 & 229.1 & $+$41.6  &~~~BAADE    & IMACS & 3976-6357;6485-10053 & 9 & 600\\
24.0 Mar 2010 & 279.0 & $+$91.6  &~~~DU PONT  & WFCCD & 3700-9500  & 7 & 900  \\
\hline
\end{tabular}
\begin{tablenotes}
\item [a] Days relative to $B$-band maximum.
\end{tablenotes}
\caption{Journal of spectroscopic observations}
\end{threeparttable}
\label{journal}
\end{table*}

\section{ANALYSIS}
\label{results}

\subsection{Reddening}

According to the \citet{sch98} IR dust maps, the Galactic reddening component 
in the direction of SN~2009mg is  $E(B-V)_{MW} =$ 0.045 mag.
In order to estimate the colour excess attributed to dust in the host 
galaxy of SN~2009mg we turn to two methods. 

\begin{figure}
\includegraphics[angle=-90,scale=0.35]{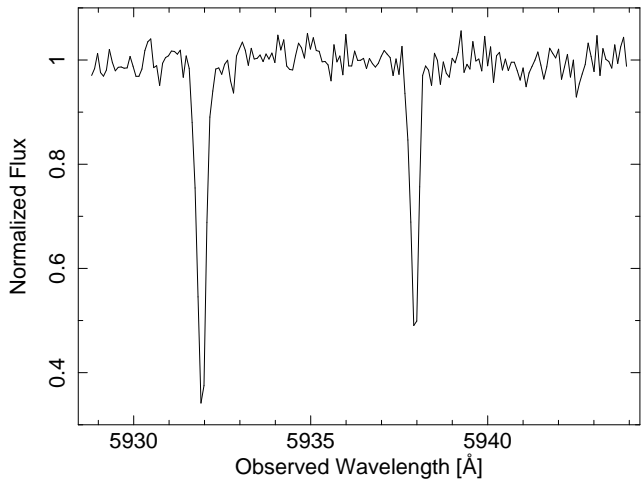}
\caption{The high resolution CLAY-MIKE spectrum displaying the Na I D1 $\lambda$5931.9 and Na I D2 $\lambda$5937.9 lines.}
\label{mike}
\end{figure}

First, from close examination of our high-resolution MIKE spectrum 
obtained on 24.3 December 2009 UT (55189.24 MJD) we identify
Na I D1 $\lambda$5931.9 and Na I D2 $\lambda$5937.9, shown in Fig. \ref{mike}, 
at the the redshift of the host galaxy. 
From this we measure an equivalent width (EW) for Na I D1 of 0.21~\AA.
Following the \citet{mun97} relation between $E(B-V)$ and EW of Na I D1 
implies a host galaxy colour excess of $E(B-V)_{host} =$ 0.07$\pm$0.02 mag. 

We have also estimated the host galaxy colour excess by determining the offset 
of the $b-v$ colour curve of SN~2009mg with respect to a 
template colour curve (Drout et al. 2011; Stritzinger et al., in preparation). 
This template colour curve was constructed from a sample of six 
unreddened SNe~Ib/c and SNe~IIb which were observed by the {\em Carnegie Supernova Project} 
(Stritzinger et al., in preparation). 
From this method we estimate an $E(B-V)_{host} = 0.18\pm0.04$ mag. This estimate is 
about a factor of 2 higher than obtained from the Na I D1 lines. 

In what follows we adopt the weighted mean of our two host colour excesses
estimates, corresponding to an $E(B-V)_{host} =$ 0.09$\pm$0.02 mag. When combined
with the Galactic colour excess, we obtain a total colour excesses of 
$E(B-V)_{tot} =$ 0.14$\pm$0.02 mag. For an $R_V = 3.1$ this corresponds to 
an $A_V =$ 0.43 mag.

\subsection{Photometric Evolution}
\subsubsection{Results}
Our broad-band photometric observations of SN~2009mg, which began 3 days after discovery, 
are plotted in Fig.~\ref{UVOT_lc}, and extend over a duration of 47 days. There are detections in all bands, 
but there are fewer detections in the three UV bands compared to the three optical bands. The $uvm2$ band 
has only one detection from summed observations. The particular brightness of the $uvw2$ band, relative 
to the $uvm2$ band, is likely to be due to red leak in the former \citep{bro10}. Comparing the peak 
colours of this SN, which are derived below, with the colours of a number of sources \citep[see their Table 12]{bro10}, 
we find most similarity with a 4000 K blackbody source. This blackbody source required red-leak corrections of 
$\sim -2.25$ mag for $uvw2$ and $\sim-0.55$ mag for $uvw1$.
 
Observations of the SN light curve began 11 days prior to $B$-band maximum. The rise to maximum is best 
observed in the $ubv$-band light curves, which show a tendency for the SN to rise faster and peak earlier in 
the bluest filters. After reaching maximum, the light curves decay at different rates depending on wavelength 
with the quickest decays observed in the bluest filters. This is consistent with both the observations of 
most SNe~Ib/c and SNe~IIb, including SNe~1993J and 2008ax \citep{sch93,lew94,ric94,pas08,rom09,tau11}, 
and with theoretical predictions \citep{arn82}.

For each filter, the observed peak magnitude ($\rm m_{peak}$) and peak time ($\rm t_{peak}$), 
listed in Table~\ref{UVOT_mag}, were determined using an average of $10^5$ Monte Carlo simulations of a cubic 
spline fit to the light curves. The errors for these parameters were taken to be the standard 
deviation of the simulated distributions. The Monte Carlo simulations were performed on each 
filter using the first 12 observations; upper limits were not used. Since {\it uvm2} has only 
a single detection, we are not able to use a cubic spline fit to determine $\rm t_{peak}$ and $\rm m_{peak}$ 
and we therefore do not provide values of these parameters for this filter. To verify our estimates of $\rm t_{peak}$ and 
$\rm m_{peak}$ for the other filters, we repeated the fitting using a simple polynomial and compared 
the results to those values derived using the Monte Carlo cubic spline fitting. The values of $\rm m_{peak}$ 
estimated from the polynomial fits are consistent with cubic spline fits to within 1$\sigma$ for the 
optical bandpasses and 2$\sigma$ for the UV bandpasses. For values of $\rm t_{peak}$ the polynomial 
fits are consistent with the cubic spline fits, in all filters, to within 1$\sigma$. The determination 
of the peak times of the separate bands (see Table~\ref{UVOT_mag}) confirms our earlier observation that 
the bluest filters peak the earliest. 

For each filter, we also calculated the absolute peak magnitude ($\rm M_{peak}$) using the observed peak 
magnitude for each filter and the distance modulus. The resulting values are given in Table \ref{UVOT_mag} 
and have been corrected for extinction and K corrected. To determine the K corrections and extinction 
corrections, we de-reddened and redshifted to the restframe a template spectrum from a similar 
Type IIb SN \citep[SN 1993J;][]{jef94}. The template spectrum of the Type IIb SN, 1993J, was used 
because the spectral range covers the full bandpass of the UVOT (i.e 1600\AA - 8000\AA). To the 
spectral template we applied the $E(B-V)$ values corresponding to the Galactic and host extinction 
using the Milky Way and Small Magellanic Cloud extinction (SMC) laws \citep{car89,pei92}, 
respectively. The extinction corrections were computed via the subtraction of synthetic magnitudes of 
an unreddened and reddened template spectrum in the observed frame. The K corrections were computed 
through the subtraction of synthetic magnitudes from the unreddened template spectrum in the rest 
and observed frames. The resulting extinction and K correction factors are provided in Table \ref{kcorr}. 
Correcting for extinction and K correcting gives a value for the absolute peak $v$ 
magnitude of SN~2009mg of $-17.68\pm0.48$, which is consistent with other SNe~IIb \citep{ric06,dro11}, 
including SN~1993J ($-17.57\pm0.24$) and SN~2008ax ($-17.61\pm0.43$) \citep{tau11}.

\begin{table}
\begin{tabular*}{9cm}{lcccccc}
\hline
UVOT Filter & {\it v}  & {\it b} & {\it u} & {\it uvw1} & {\it uvm2} & {\it uvw2} \\
\hline
\hline
Extinction &$-$0.42 &$-$0.55 &$-$0.67 &$-$0.81 &$-$1.09 & $-$0.99\\
K correction &0.00 &$-$0.01 &$-$0.05 & $-$0.04& $-$0.04& $-$0.03\\\end{tabular*}
\caption{Total extinction and K correction factors for the six UVOT bandpasses in magnitudes.}
\label{kcorr}
\end{table}

Plotted in the middle panel of Fig.~\ref{UVOT_lc} are the $(u-b)$ and $(b-v)$ colour curves. From the 
start of observations, with $(u-b)\sim0.84$ mag and $(b-v)\sim0.37$ mag, the colour curves are observed 
to evolve from the red to the blue until the time of photometric maximum, at which point they reach 
minimum values of 0.52 mag and 0.14 mag for $(u-b)$ and $(b-v)$, respectively. After photometric 
maximum both colour curves evolve towards the red, reaching $\sim$ 1 mag, and then cease evolving. 

The spectra for SN~2009mg have spectral range covering the UVOT $v$ bandpass, 
we were therefore able to compute optical $v$-band K corrections. The results are plotted in 
the lower panel of Fig.~\ref{UVOT_lc} at epochs corresponding to when spectroscopic 
observations were obtained. The evolution of the K correction for the $v$ filter tends 
to follow the $b-v$ colour curve evolution. Interestingly, this behaviour mimics what is observed in 
thermonuclear SNe \citep{nug02}.

\subsubsection{Comparison with other SNe~IIb}

\begin{figure*}
\includegraphics[angle=-90,scale=0.50]{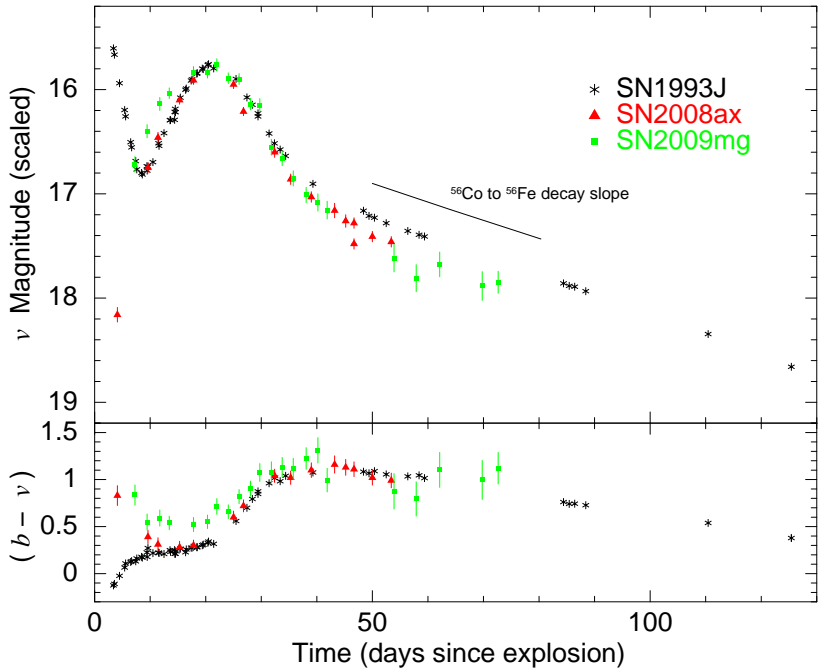}
\caption{Comparison of the $v$-band light curves of SN~2009mg to those of the prototypical Type~IIb SNe~2008ax and 1993J.
The {\em top panel} shows the $v$-band light curve of SN~2009mg compared with those of SNe~1993J and 2008ax.
As the explosion date of SN~2009mg is unknown, the peak time as been shifted to match that of SN~2008ax.
The lower panel shows the ($b-v$) colour evolution for each of these objects, corrected for extinction.}
\label{comparison}
\end{figure*}

The brightening to maximum and the subsequent evolution of SN~2009mg's optical light curves
follows the classic behavior of a normal SN~IIb. In the top panel of Fig.~\ref{comparison} 
we compare the $v$-band light curve of SN~2009mg to those of SNe~1993J \citep{lew94} and 2008ax \citep{rom09}. 
Note no extinction corrections have been applied to the photometry and cosmological corrections 
are ignored since the redshift of each object is low. Here the light curves of SNe~1993J and 
2008ax have been scaled to the peak brightness of SN~2009mg. Since the time post-explosion 
is uncertain for SN~2009mg, the time for this SN has been scaled to the well-constrained 
explosion date of SN~2008ax. Scaling the time axis of SN~2009mg to that of SN~2008ax allows 
us to estimate the explosion epoch to be $22\pm2$ days prior to $v$-band maximum, which 
corresponds to 3 December 2009.

Figure~\ref{comparison} reveals that the $v$-band light curve of each SN is strikingly similar, 
particularly from maximum light and onwards. Prior to maximum, the $v$-band light curve of SN~2009mg 
appears to be slightly broader than the comparison objects. The broader rise for SN~2009mg
may be an indication that the ejecta mass of SN~2009mg is greater than for SNe~1993J and 2008ax,
or the ejecta velocity of SN~2009mg is lower than the comparison objects.

The largest difference between the three objects is observed at the earliest epochs. In SN~1993J there is an 
initial drop in luminosity, which is related to a cooling phase that follows shock breakout, while 
for SN~2008ax there is a marginal detection of such a decline in the $u$ band \citep{rom09}. 
No post-shock breakout cooling phase is documented in the $v$ light curve of SN~2009mg, nor at 
bluer wavelengths (see Fig.~\ref{UVOT_lc}). This is a strong indication that the observations 
of SN~2009mg did not commence early enough to catch the initial upturn in luminosity and may 
also be an indication that the explosion date for SN~2009mg is earlier than the 22 days estimated 
from comparison with the light curve of SN~2008ax.

From $\sim$10 days post-explosion, the light curves of all three objects exhibit a similar overall 
behaviour, a rise to a peak, followed by an exponential decay, which then flattens to a shallower 
linear decay from $\sim 40$~days. After the peak, they all appear to decay at the same rate. 
Given the comparison objects are nearly identical to SN~2009mg from maximum light onwards 
we attempt to quantify their similarities by measuring 
the  $\Delta m_{15}$ diagnostic for the $b$- and $v$-band light curves.  
Here $\Delta m_{15}$ is taken to be the magnitude difference between the peak SN 
brightness and its brightness measured 15 days post-peak  \citep{phi93}.

For the $v$ light curves we obtain $\Delta m_{15}$($v$) values of $0.948\pm0.004$ mag, 
$1.19\pm0.07$ mag and $1.17\pm0.08$ mag for SNe~1993J, 2008ax and 2009mg, respectively. 
These values of $\Delta m_{15}$($v$) are consistent with, but towards the steeper end, of the range of values 
determined for a sample of SNe~Ib/c $V$-band light curves \citep{dro11}, where the mean $\Delta m_{15}$($V$) 
for their sample is $0.87\pm0.25$ mag. However, $\Delta m_{15}$($v$) for SN~2008ax is slightly 
faster than the $V$-band decline rate found by \cite{tau11} and \cite{dro11}. For the $b$-band light 
curve, $\Delta m_{15}$($b$) for the three objects is: $1.67\pm0.01$ mag (SN~1993J), 
$1.57\pm0.06$ mag (SN~2008ax) and $1.46\pm0.09$ mag (SN~2009mg). Note the value of 
$\Delta m_{15}$($b$) for SN2008ax is consistent with the value of $\Delta m_{15}$($B$) obtained by \citet{tau11}. 
SN~2009mg declines more slowly in both the $v$ and $b$ bands in comparison to 1993J, while
it is more consistent with the evolution of SN~2008ax. 

The lower panel of Fig.~\ref{comparison} displays the $(b-v)$ evolution for SNe~1993J, 2008ax and 2009mg.
We have applied extinction corrections, with values of $A_v=0.94$ and $A_b=1.24$ for SN~2008ax \citep{rom09}, 
and, $A_V=0.58$ and $A_B=0.78$ for SN~1993J \citep{lew94}. Although we have not corrected for any systematic 
offset between the colour curves of each object due to the different photometric systems, these 
differences are expected to be minimal with ($B-V$) $\sim$ ($b-v$) $\sim$ 0.01 mag \citep{poole}. From 
20 days post-explosion, the $(b-v)$ colour curves exhibit a similar evolution for all three SNe~IIb. During 
the initial 20 days post explosion, the colour evolution of SN~2009mg is most similar to SN~2008ax. 
Observations of SN~2009mg commenced at 7 days post-explosion at which ($b-v$) for SN~2009mg and 
SN~2008ax is 0.83 and 0.56, respectively. From this point the colour curves of SNe~2009mg and 
2008ax are observed to dip towards the blue, reaching minimum values of ($b-v$) of $\sim0.52$ 
and $\sim0.28$. The colour curves then evolve back towards the red, finally leveling 
off at ($b-v) \sim 1$ before observations ceased. The behaviour of ($b-v$) colour curve for SNe~2008ax 
and 2009mg appear to be a common feature for Type Ib/c and Type IIb SNe
\citep[e.g.][]{rom09,str09,dro11}.

To constrain and compare the late-time temporal behaviour, we computed the decay rate during the shallow 
decline phase, from 40 days, for the $u$, $b$ and $v$ light curves of  SN~2009mg, and for the $V$-band 
light curve of 1993J. The decay rate was computed using a linear least squares fit. As the observations of SN~2009mg ceased at $\sim70$ days
post explosion, we fit the linear function to the $u$, $b$ and $v$ light curves from 40 days post explosion to
the end of observations at $\sim70$ days. This resulted in decay rates of $-0.004\pm0.011$ mag day$^{-1}$,
$0.022\pm0.004$ mag day$^{-1}$ and $0.025\pm0.003$ mag day$^{-1}$ for the $u$, $b$ and $v$ light curves, respectively.
The decay rate of SN~1993J in the $V$ band during the same period is $0.0213\pm0.0002$ mag day$^{-1}$, which is
consistent with the decay rate of the $v$ filter of SN~2009mg. The $v$ decay rate of SN~2009mg is inconsistent, 
at 3$\sigma$ confidence, with the decay rate of $^{56}$Co to $^{56}$Fe, which is 0.0097 mag day$^{-1}$. The $u$ band light curve of SN~2009mg 
appears to cease decaying at late times, with the decay rate consistent with being constant. Similar behaviour is
also observed in the $u$-band light curves of the Type Ib/IIb SNe~2007Y and 2008aq, which are observed not to 
decay at late times, but to increase in brightness between $\sim 20$ and $\sim90$ days post-peak \citep{str09}. 
The $u$ light curve of SN~2008aq is observed to decrease in brightness again after $\sim90$ days. The $U$-band light curve
of SN~1993J was also reported to remain almost constant from $\sim 50$ to $\sim 125$ days past maximum \citep{lew94}.

\subsection{Optical Spectroscopy}

\begin{figure}
\includegraphics[angle=0,scale=0.64]{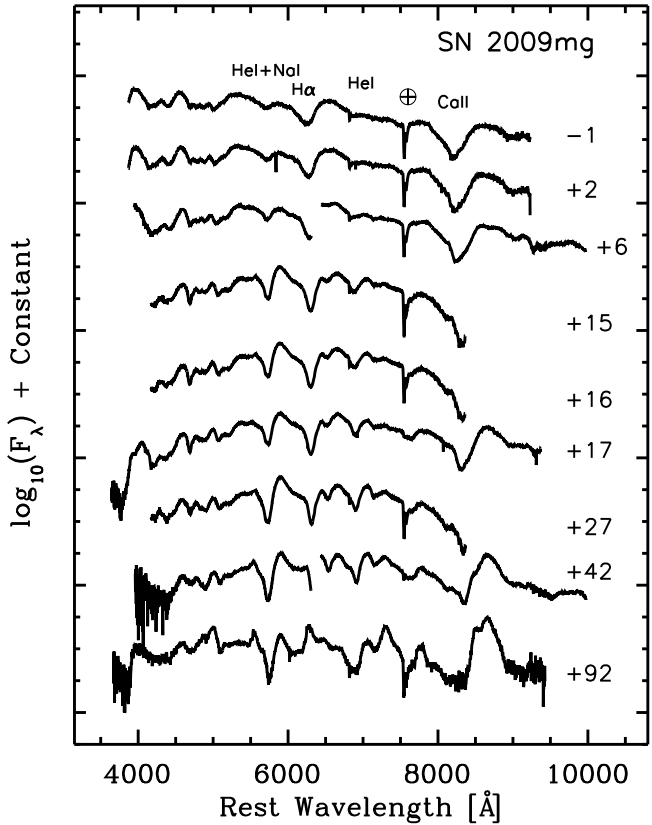}
\caption{Spectroscopic sequence of SN~2009mg ranging from $-$1 to $+$92 days 
relative to $B$-band maximum. Each spectrum has been corrected to the rest frame of the host galaxy 
adopting the redshift $z = 0.0076$, and for presentation purposes shifted in 
flux by an arbitrary constant.}
\label{spect_evol}
\end{figure}

The spectroscopic sequence of SN~2009mg, shown in Fig.~\ref{spect_evol}, reveals
the typical evolution of a SN~IIb. At early phases H$\alpha$, and absorption features
associated with Fe~II and Ca~II dominate the spectrum. In addition, during the earliest 
observed epochs features attributed to He I and Na I are also discernible, and over time,
grow in strength. As is common for SNe IIb, the strength of the He I lines compared to 
H$\alpha$ increased as the SN evolved. In the case of SN~2009mg, the relative strength does not
appear to grow as quickly as in SN~2008ax. For SN~2008ax, H$\alpha$ and He I $\lambda$5876 were of
comparable strength $\sim$30 days after the explosion, while the H$\alpha$ absorption was negligible 
by $+$56 days post explosion \citep{pas08}. For SN~2009mg, assuming an explosion date of 3 December 2009, 
the signal to noise ratio of these two features is comparable in strength by day 45 post explosion. 
Our series of spectra do not allow us to place confirm constraints on the epoch in which H$\alpha$ 
completely disappears, however, H$\alpha$ is still strong 60 days after the explosion, and is
absent in our next spectrum taken on 110 days post-explosion. 

\begin{figure}
\includegraphics[angle=0,scale=0.54]{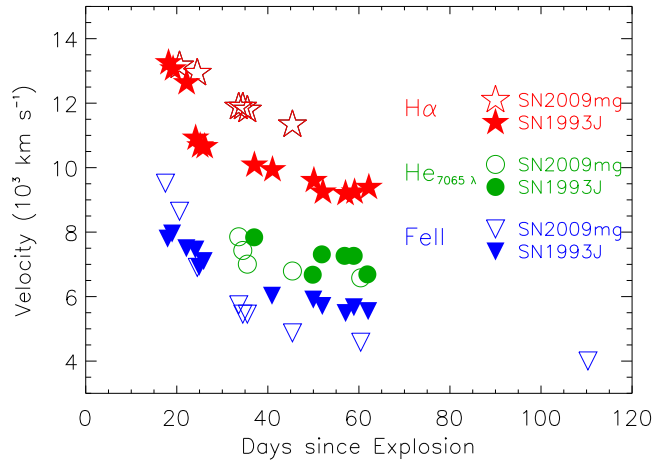}
\caption{Time evolution of line velocities for H$\alpha$, He~I $\lambda$7065 and Fe~II $\lambda$5169.
For comparison the evolution of same line velocities for SN~1993J are plotted 
\citep[][]{pas08}.}
\label{expan_comp}
\end{figure}

In Fig.~\ref{expan_comp} we plot the time evolution of the blueshifts of the absorption 
features due to H$\alpha$, Fe~II $\lambda$5169 and He~I $\lambda$7065. Included in this 
figure are also the time evolution of these lines determined from the spectroscopic sequence 
of SN~1993J \citep[][]{pas08}. These measurements were determined from the minima of the 
P-Cygni features. The evolution of the He I line  is similar in both objects, while for 
H$\alpha$, the velocity is similar at the start of the observations, but unlike SN~1993J, 
the strength of this feature in SN~2009mg does not drop steeply after day 25, but remains 
consistently higher. The Fe II line velocity of SN~1993J is found to be higher at the start 
of observations in comparison to SN~2009mg, but declines at a faster rate. It is interesting 
to compare Fig.~\ref{expan_comp} with the equivalent figure for SN~2008ax in 
\citet[][see their Fig.~6]{pas08}. The evolution of the line velocities of H$\alpha$, 
He I $\lambda$7065, and Fe II $\lambda$5169 for both SNe 2008ax and SN~2009mg behave very 
similarly, although the H$\alpha$ line in SN~2008ax remains at a slightly higher velocity 
at late times compared to SN~2009mg and Fe II $\lambda$5169 drops off at slightly a faster 
rate for SN~2008ax. This may suggest that the H envelope mass and the blastwave evolution 
are similar for the two SN.

\subsection{X-ray observations and mass-loss rate}
\label{XRTresults}

Analysis of the X-ray observations show no significant detection for SN~2009mg. Using 
a standard ten-pixel radius (24$\farcs$5) source region and extracting the background 
from a local, source-free region to account for sky background 
and for diffuse emission from the host galaxy, the 3$\sigma$ upper 
limit to the XRT net count rate is $5.3\times 10^{-4}\; {\rm cts\;s^{-1}}$.
This corresponds to an unabsorbed (0.2 $-$ 10 keV band) X-ray flux of 
$<2.5\times 10^{-14}\; {\rm erg\;cm^{-1}}\;{\rm s^{-1}}$,
and a luminosity of $<5.0\times 10^{39}\;\rm {erg\;s^{-1}}$ for an adopted 
thermal plasma spectrum with a temperature of kT~=~10~keV, a Galactic foreground 
column density of $N_H = 3.5\times10^{20}\;\rm {cm^{-2}}$ \citep{dic90},
and a distance of 33 Mpc. The upper limit derived for the luminosity is consistent
with values determined for other stripped core-collapse SNe including 
SN~2008ax \citep{imm06,imm07,rom09}.

In core-collapse SNe the likely cause of X-ray emission is the interaction of the SN blast wave with circumstellar material
\citep{che82,imm01,imm07}. We use the X-ray non-detection to place an upper limit on the mass-loss
rate of the progenitor. Following the methodology described in \cite{imm07}, we derive a mass-loss rate of
$<1.5\times 10^{-5}\;M_\odot\;\rm{yr^{-1}} (\nu_w/10\;{\rm km\,s^{-1}})$, assuming the speed of the
blast wave as $\nu_s=10,000\;{\rm km\;s^{-1}}$ and scaled for a stellar wind speed of $\nu_w=10\; {\rm km\;s^{-1}}$.
The velocity of the Fe~II $\lambda$5169 feature is representative of the velocity of the photosphere \citep{des05}.
Therefore, we are able to assume a value of $10,000\;{\rm km\;s^{-1}}$ for the blast wave velocity, since
the peak velocity of Fe II $\lambda$5169 feature is approximately $10,000\;{\rm km\;s^{-1}}$ at 
maximum light. The blast wave velocity is proportional to the mass-loss
rate and so assuming a blast wave velocity of $10,000\;{\rm km\;s^{-1}}$, rather than using an average value
from across the duration of the XRT observations, gives us a conservative upper limit to the mass-loss rate.
The resulting $3\sigma$ upper limit lies within the range of the mass-loss rate determined for SN~1993J
\citep[$10^{-5}-10^{-4}$;][]{imm07}, and is also consistent with SN~2008ax
\citep[$(9\pm3)\times 10^{-6}$;][]{rom09} and Type IIP SNe \citep[$10^{-6}-10^{-5}$;][]{che06,imm07}, 
which have also been shown to have low mass-loss rates.

\section{Discussion}
\label{discussion}

As we have shown in the previous section, the observation properties of SN~2009mg closely resemble 
that of other normal SNe~IIb. We now endeavour to place limits on the explosion properties of this object. 

\begin{figure}
\includegraphics[angle=90,scale=0.34]{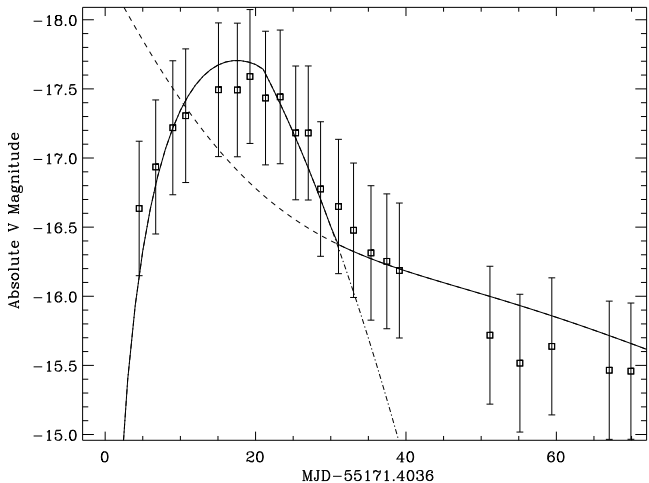}
\caption{The best-fit to the $v$-band light curve following the methodology of \protect\cite{ric06}. The dot-dashed
line shows the fit using the Arnett model, while the dotted line show the fit of the Jeffery model. The
overall combined best fit model is given by the solid line. }
\label{model_lc}
\end{figure}

Using the method described by \cite{ric06}, we proceed to model the $v$-band light curve of SN~2009mg
to provide reasonable estimates on the kinetic energy ($E_k$), ejected mass ($M_{ej}$), and ejected nickel mass 
($M_{Ni}$). This simple analytic method uses the model of \cite{arn82} to describe the peak of the light curve
while the linear tail is best represented by the model of \cite{jef99}. To constrain these three parameters,
a Monte Carlo simulation of the $v$-band light curve of SN~2009mg was computed with $10^6$ realizations.
This analysis is best performed using the UVOIR light curve, but there are no IR observations of this SN to our knowledge.
Since the $v$-band light curve is a reasonable proxy for the bolometric light curve, we use this as a substitute to
the UVOIR light curve and apply a Bolometric correction of $-$1.48 mag, as described in \cite{ric06}. In this 
simulation the routine switches from the Arnett model to the Jeffery model at day 31 post-explosion. The best
$\chi^2$ fit ($\chi^2/D.O.F=1.1$ ) results in $E_k = 0.15^{+0.02}_{-0.13}\times10^{51}$~erg, 
$M_{ej} = 0.56^{+0.10}_{-0.26}M_\odot$, and $M_{Ni} = 0.10\pm0.01 M_\odot$. The best fit model is shown in
Fig. \ref{model_lc}. The resultant parameters are similar to, but lower than the values
calculated by \cite{ric06} for the Type IIb SNe 1996cb and 1993J. Since the Jeffery model assumes 
that the $^{56}$Ni has decayed and the ejecta is optically thin, it should ideally be fit 
to data between $\sim$100 and $\sim$150 days. We therefore verified our result by repeating 
the simulation, fixing the time at which the simulation transfers from the Arnett to the Jeffery 
model to be 50~days post explosion. The resulting values for $E_k$, $M_{ej}$ and $M_{Ni}$ are 
consistent with the values determined with the transfer at 31 days post explosion.

The favoured progenitors, of Type SNe~IIb, are massive stars in binary systems \citep[][see references within]{smi11,cla11}. 
Models of stripped helium stars in binary systems, are provided by \cite{shi90} and \cite{pod93}. \cite{shi90} conclude that 
smaller helium stars undergo more extensive mixing and eject smaller masses than larger helium stars 
and so form light curves with steeper tails. This is because the ejecta is more transparent when the 
ejecta mass is small and mixed \citep{shi90,imw97} and thus allows more $\gamma$-rays to escape before they can 
be thermalized. When $\gamma$-rays are fully trapped the light curves decay slowly at a rate consistent with the 
decay rate of $^{56}$Co. Therefore, one indication of the degree of mixing in the progenitor is the decay rate of 
the tail of the light curve. The decay rate of the tail of SN~2009mg in the $v$ band is $0.025\pm0.003$ mag day$^{-1}$, 
which is steeper, at $3\sigma$ confidence, than the decay rate of $^{56}$Co to $^{56}$Fe, which decays at a rate 
of 0.0097 mag day$^{-1}$ (see Fig. \ref{comparison}). This suggests that there is leakage of $\gamma$-rays 
and hence some degree of mixing of the progenitor \citep{shi90}. This would indicate that the progenitor is 
likely to be on the small side of the mass distribution. This is consistent with the light curve modeling, 
which suggests values of $E_k$, $M_{ej}$ and $M_{Ni}$ similar to, but lower than those of 1993J and 1996cb. 

\section{Conclusions}
\label{conclusions}
SN~2009mg appears to be a normal SN~IIb and exhibits properties similar to other normal, well-observed SNe~IIb.  
Modelling the $v$-band light curve, we find best fit parameters for kinetic energy ($E_k$) of 
$0.15^{+0.02}_{-0.13}\times 10^{51}\;{\rm erg}$, an ejecta mass ($M_{ej}$) of $0.56^{+0.10}_{-0.26}M_\odot$, and a nickel 
mass ($M_{Ni}$) of $0.10\pm0.01M_\odot$.

The decline rate of the light curve tail starting from 40 days past $B$-band maximum is inconsistent, at 
$3\sigma$ confidence, with the decline rate of $^{56}$Co. This indicates that there is leakage of $\gamma$-rays 
out of the ejecta and suggests the progenitor star was on the lower end of the stellar mass distribution.

\section{Acknowledgments}
We thank the referee for their useful comments. We also thank N. Morrell for performing spectroscopic 
observations. This research has made use of data obtained from the High Energy Astrophysics 
Science Archive Research Center (HEASARC) and the Leicester Database and Archive Service 
(LEDAS), provided by NASA's Goddard Space Flight Center and the Department of Physics 
and Astronomy, Leicester University, UK, respectively. S.R.O., A.A.B. and N.P.M.K. acknowledge the support of 
the UK Space Agency. J.L.P. acknowledges support from NASA through Hubble Fellowship 
Grant HF-51261.01-A awarded by STScI, which is operated by AURA, Inc. for NASA, under 
contract NAS 5-2655. M.H. acknowledges support by the Millennium Center for Supernova Science (P10-064-F).


\bibliographystyle{mn2e}   
\bibliography{SN2009mg} 

\IfFileExists{\jobname.bbl}{}
 {\typeout{}
  \typeout{******************************************}
  \typeout{** Please run "bibtex \jobname" to optain}
  \typeout{** the bibliography and then re-run LaTeX}
  \typeout{** twice to fix the references!}
  \typeout{******************************************}
  \typeout{}
 }

\end{document}